\documentclass[11pt,twoside,a4paper,cr,final]{cms-tdr}
\def\svnVersion{194012M}

\begin{document}

\hyphenation{had-ron-i-za-tion}
\hyphenation{cal-or-i-me-ter}
\hyphenation{de-vices}

\RCS$Revision: 187380 $
\RCS$HeadURL: svn+ssh://svn.cern.ch/reps/tdr2/papers/XXX-08-000/trunk/XXX-08-000.tex $
\RCS$Id: XXX-08-000.tex 187380 2013-05-24 13:11:41Z alverson $
\newlength\cmsFigWidth
\ifthenelse{\boolean{cms@external}}{\setlength\cmsFigWidth{0.85\columnwidth}}{\setlength\cmsFigWidth{0.4\textwidth}}
\ifthenelse{\boolean{cms@external}}{\providecommand{\cmsLeft}{top}}{\providecommand{\cmsLeft}{left}}
\ifthenelse{\boolean{cms@external}}{\providecommand{\cmsRight}{bottom}}{\providecommand{\cmsRight}{right}}
\cmsNoteHeader{CR-2013/155} 
\title{Searches for new physics with leptons and jets at CMS}

\address[osu]{The Ohio State University}
\author{Carl Vuosalo on behalf of the CMS Collaboration}

\date{\today}

\abstract{
A variety of models of physics beyond the standard model predict new particles that decay to 
leptons, jets, or both together. These 
models include axigluons, colorons, diquarks, excited quarks, heavy long-lived charged particles,
leptoquarks, Randall-Sundrum gravitons,
string resonances, and new vector bosons (right-handed W and Z').
Using the data collected in 2011 and 2012 at center-of-mass energies of 7 and 8~TeV,
the CMS collaboration has
performed searches for these new particles in channels with leptons and jets. The results of
these searches will be presented. No evidence of new physics has been observed,
and these results set new limits on the parameters of these models.
}

\hypersetup{%
pdfauthor={CMS Collaboration},%
pdftitle={Searches for new physics with leptons and jets at CMS},%
pdfsubject={CMS},%
pdfkeywords={CMS, physics}}

\maketitle 

\section{Introduction}
\label{sec:intro}

A major goal of the Compact Muon Solenoid (CMS) experiment~\cite{cmsdet} at the Large Hadron Collider
is the discovery of new physics (NP) beyond the standard model (SM).
Many NP theoretical models predict
production of resonances that decay to paired jets or to leptons and jets. During 2012, the CMS
detector recorded 19.6~fb$^{-1}$ of pp data at a center-of-mass energy of 8~TeV. Several CMS
analyses exploited this rich data set to search for these NP signals.

\section{Search for a Heavy Neutrino and $\textrm{W}_{\textrm{R}}$}
\label{sec:W_R}
A proposed, new right-handed $\textrm{SU}_{\textrm{R}}(2)$ symmetry group would produce
new gauge bosons, called $\textrm{W}_{\textrm{R}}$ and Z', and three heavy, right-handed
neutrinos. Production of a $\textrm{W}_{\textrm{R}}$ resonance would produce two leptons and two jets:

\begin{equation}
W_R \rightarrow \ell_1 N_\ell \rightarrow \ell_1 \ell_2 W_R^* \rightarrow \ell_1 \ell_2 q q' \rightarrow \ell_1 \ell_2 j j.
\end{equation}

\begin{figure}[!htb]
\includegraphics[width=0.45\linewidth]{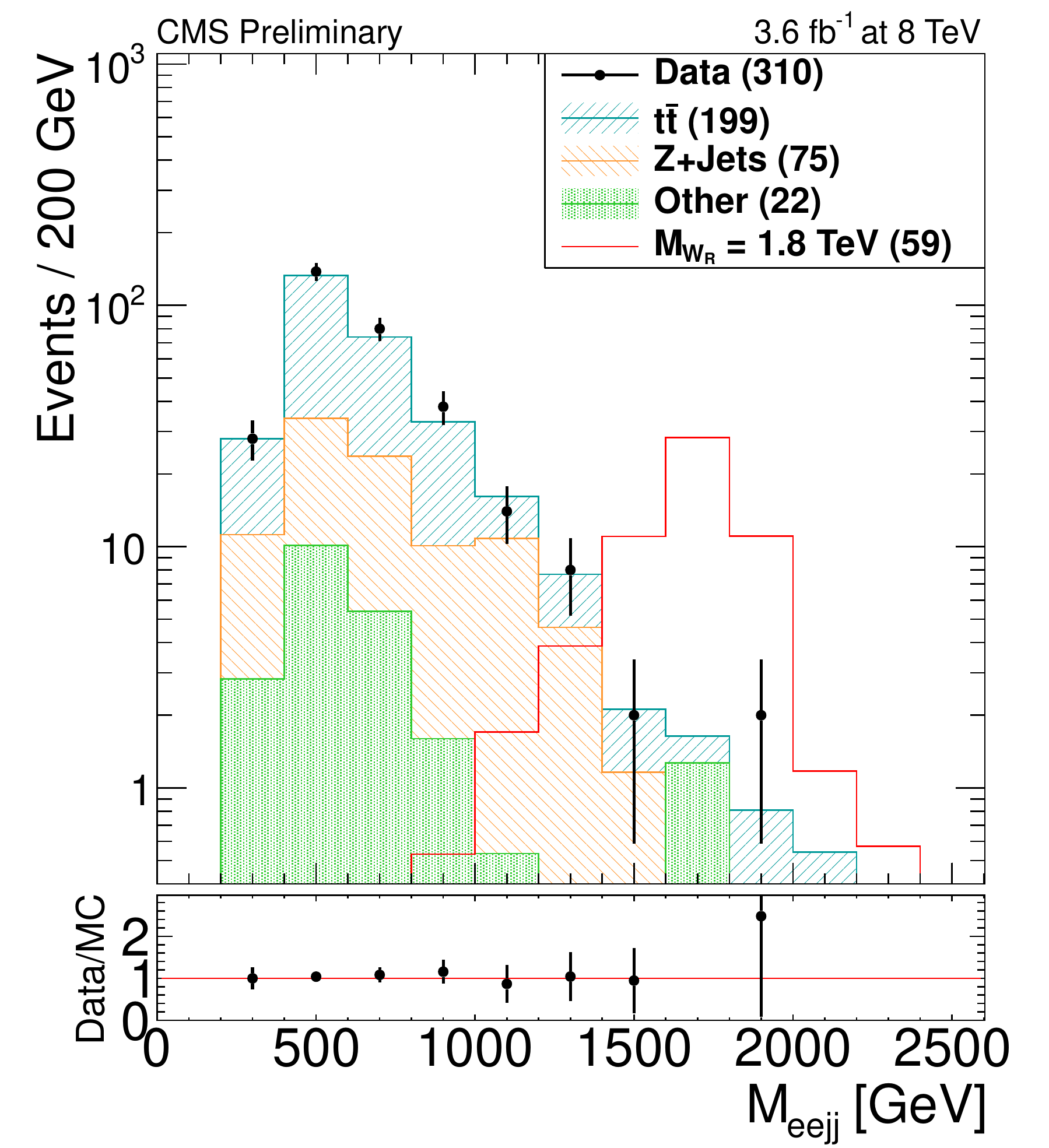}
\hfill
\includegraphics[width=0.45\linewidth]{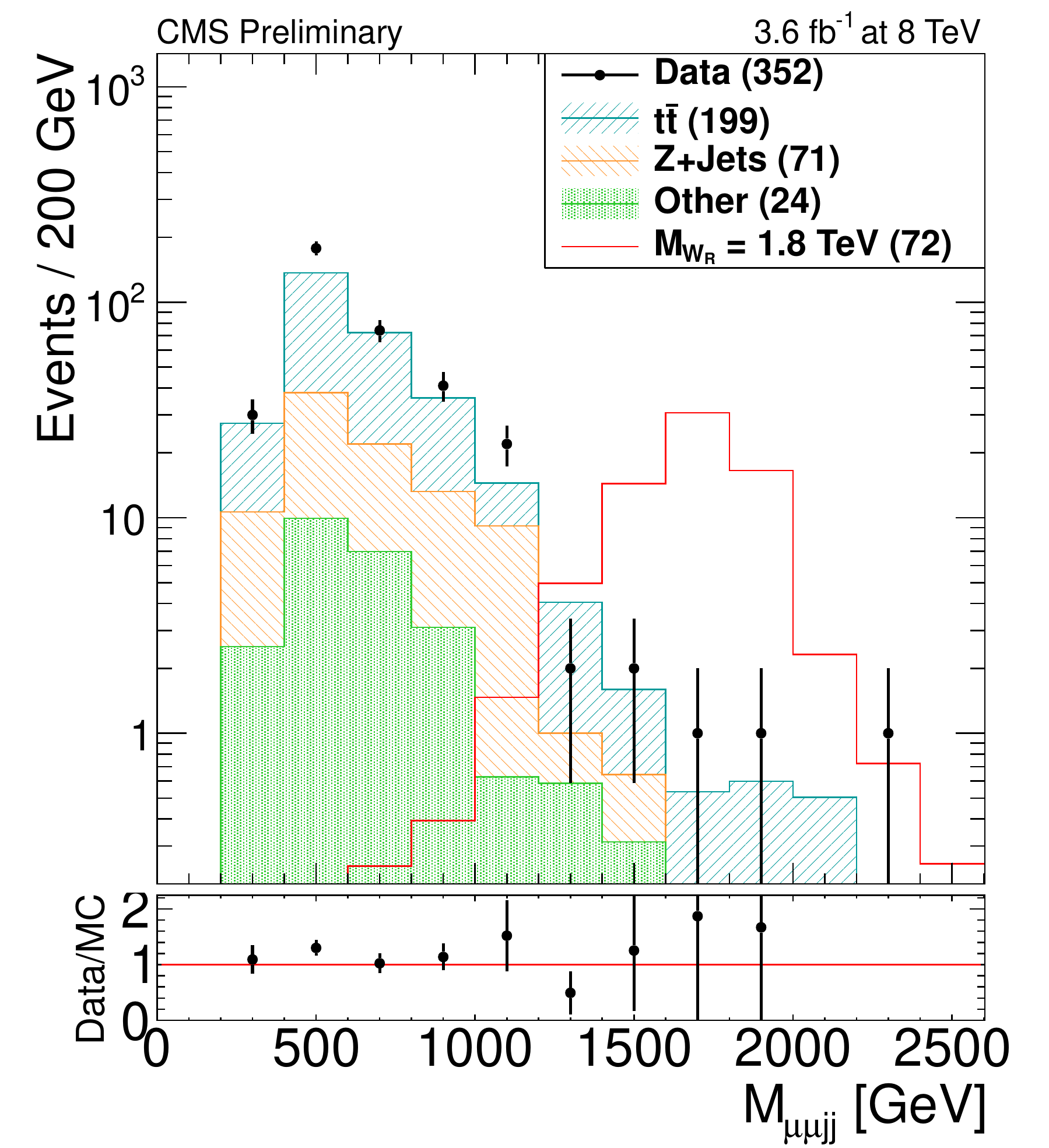}
\caption{Four-object mass distribution for $eejj$ (left) and $\mu \mu jj$ (right) 
events surviving event selection criteria using the 
collision data collected in 2012.  The multijet and other minor SM backgrounds (e.g. diboson) 
are combined into a single ``Other'' category, and the signal mass point 
$M_{W_R}=1800$~GeV, $M_{N_{\ell}}=900$~GeV, is included for comparison.  The uncertainty 
in the data/MC ratio includes the statistical uncertainty on both the reconstructed events in 
data as well as the background expectation.
\label{fig:evtsPostMll}
}
\end{figure}

CMS has performed a search for this decay with 3.6~fb$^{-1}$ of 2012 8 TeV pp
data~\cite{exo1217}. This search utilizes the $M_{\ell\ell jj}$ invariant mass distribution, in which
the $\textrm{W}_{\textrm{R}}$ should produce a clear signal peak compared to the background,
as shown in Fig.~\ref{fig:evtsPostMll}. Both electron and muon channels are included in the 
search. Background estimates were derived with data-driven techniques, except for the Z-plus-jets
contribution, which was determined from simulation.

The systematic uncertainty for the signal is 15\% and 
comes mostly from parton distribution function (PDF)
uncertainty when generating the simulated signal samples. 
The background systematic uncertainties ranges from 20--50\%, which are mostly
shape uncertainties.

\begin{figure}[!htbp]
\includegraphics[width=0.45\linewidth]{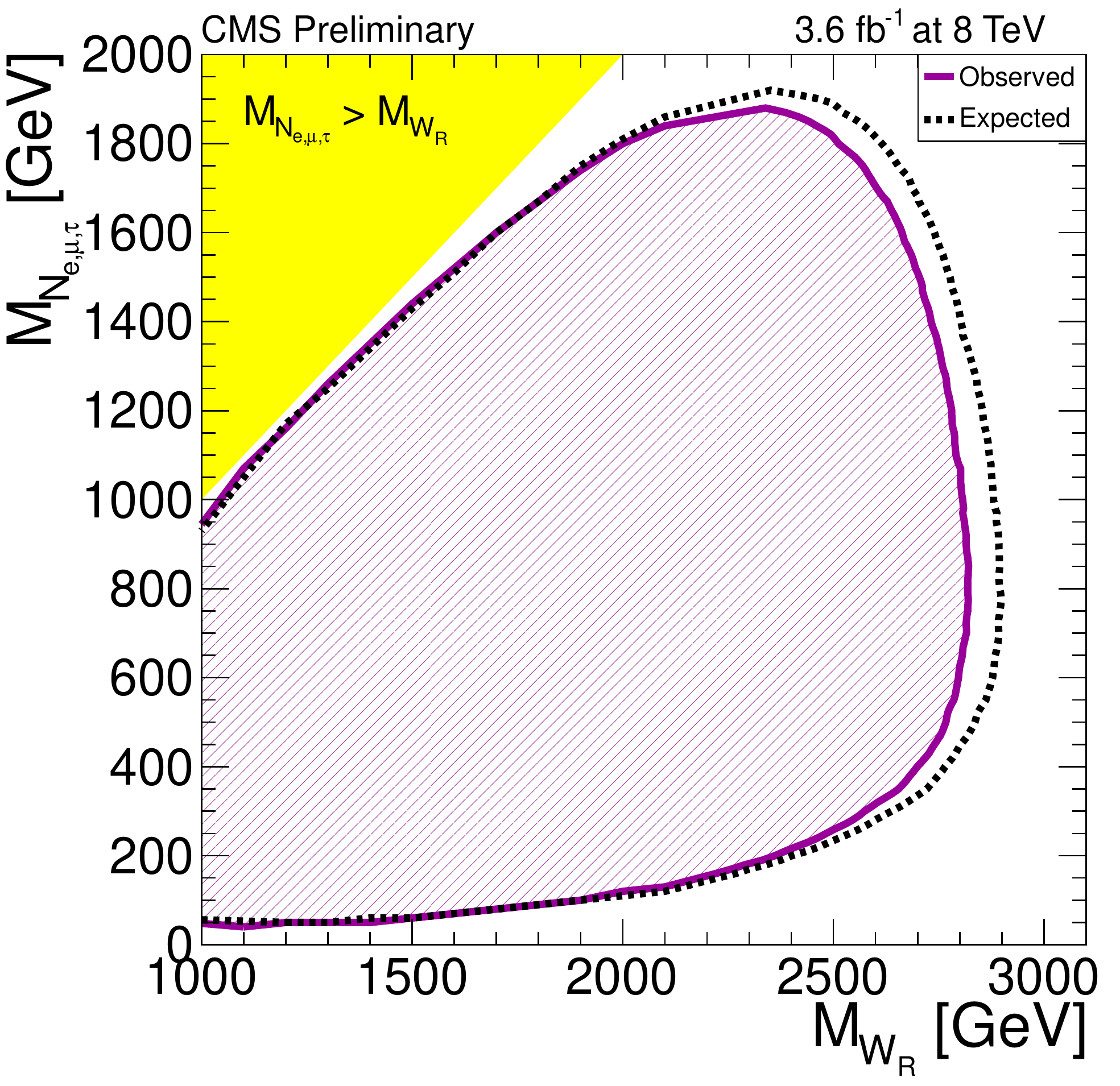}
\hfill
  \includegraphics[width=0.45\textwidth]{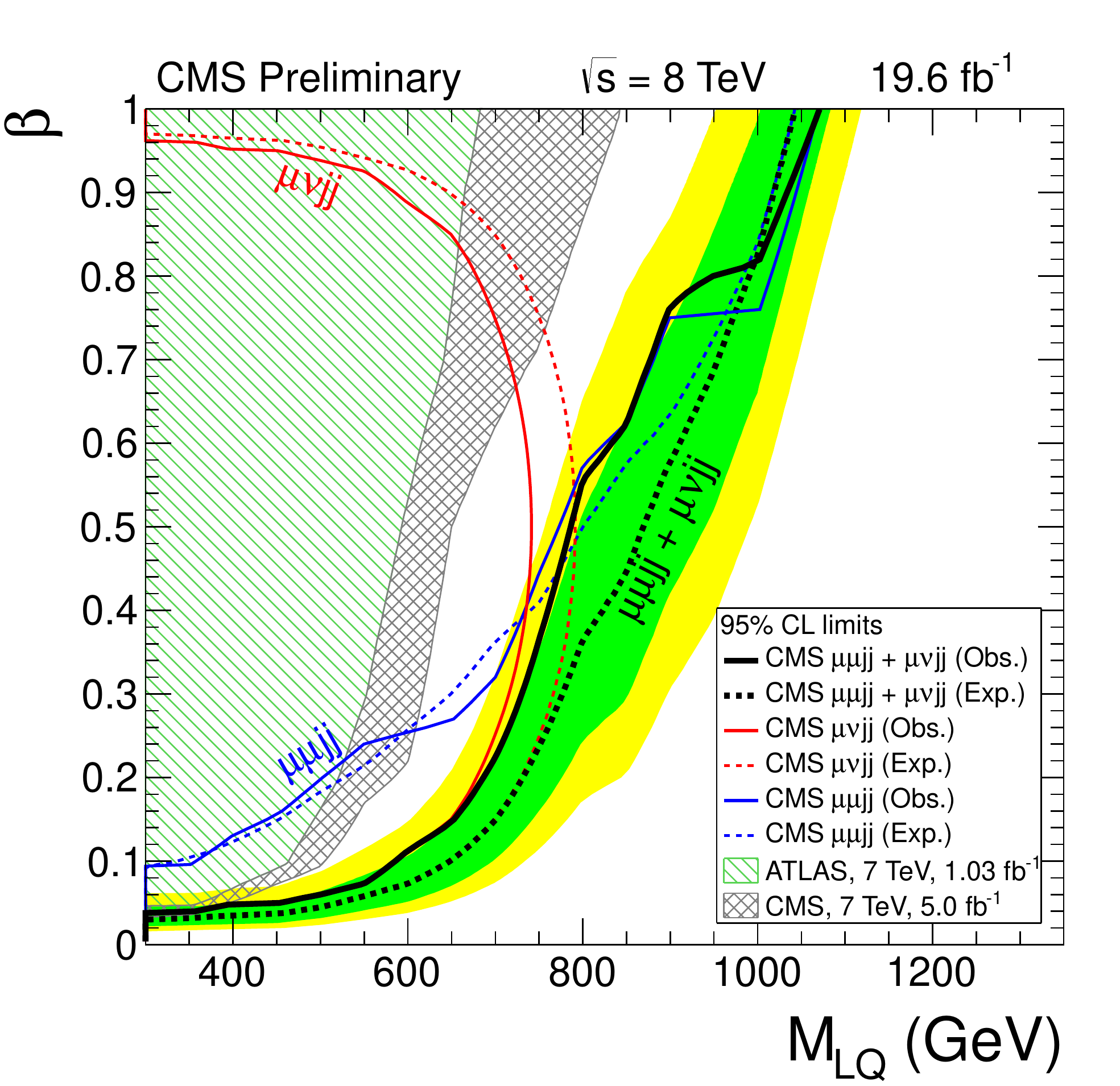}

\caption{On the left, for the $\textrm{W}_{\textrm{R}}$ search,
the 95\% confidence level exclusion region in the
$(M_{W_R},M_{N_{\ell}})$ plane  obtained from combining the electron 
and muon channels. On the right, for the leptoquark search,
the expected and observed exclusion limits at $95\%$ CL on second-generation leptoquark mass as a function of the branching fraction $\beta$.  
The dark green and light yellow expected limit uncertainty bands represent the 68\% and 95\% confidence intervals on the combination. Limits for the individual $\mu\mu jj$ and $\mu\nu jj$ channels are also given as in the $M_{\mathrm{LQ}}$ versus $\beta$ plane. Solid lines represent the observed limits in each channel, and dashed lines represent the expected limits.
  The leftmost shaded region is excluded by the most recent ATLAS 7TeV result, and the rightmost shaded region is excluded by the CMS 7 TeV Result.}
\label{fig:limitsWRemu}
\end{figure}

As can be seen in Fig.~\ref{fig:evtsPostMll}, data matches the background estimates within the
uncertainties, and no significant excess is observed. Exclusion limits were calculated from a shape
analysis of the $M_{\ell\ell jj}$ distribution. The best limits are achieved by combining the
electron and muon channels, 
with the assumption of left-right symmetry ($g_L=g_R$) and three degenerate
generations of heavy
neutrinos. These limits,  as shown on the left in Fig.~\ref{fig:limitsWRemu}, extend up to 2800 GeV on the
mass of the $\textrm{W}_{\textrm{R}}$.

\section{Search for Pair Production of Second-Generation Scalar Leptoquarks}
Leptoquarks are proposed particles with fractional charge that couple to leptons and quarks,
have color, and can be scalar or vector particles. CMS has performed a search for
pair production of
scalar leptoquarks with the full 19.6~fb$^{-1}$ of 2012 8 TeV pp data~\cite{exo1242}.
The search uses two signatures for leptoquark decay: a pair of leptoquarks each decaying
to a muon and a jet (two muons and two jets); and one leptoquark decaying to a muon and a jet,
and the second decaying to a neutrino and a jet (one muon, two jets, and missing energy).

The major backgrounds are from W plus jets and t$\bar{\textrm{t}}$ plus jets, and
background estimates were obtained through data-driven techniques. The systematic uncertainty
for signal is 5\%, mostly due to the uncertainty on the integrated luminosity measurement.
The uncertainties for backgrounds range from 14--24\% and are mostly from
jet energy resolution and muon energy scale uncertainties.

The number of observed data events matches the number of estimated background events
within the uncertainties, so limits were calculated. Combining the limits for the two channels, and
plotting them in terms of the leptoquark mass and the branching ratio of the leptoquark
decaying to a lepton and quark, gives the results shown in Fig.~\ref{fig:limitsWRemu}.
These limits extend up to 1070 GeV on the leptoquark mass.

\section{Search for Long-lived Charged Particles}
\begin{figure}[tbhp]                                                                                       
 \begin{center}
  \includegraphics[clip=true, trim=0.0cm 0cm 1.0cm 0cm, width=0.45\linewidth]{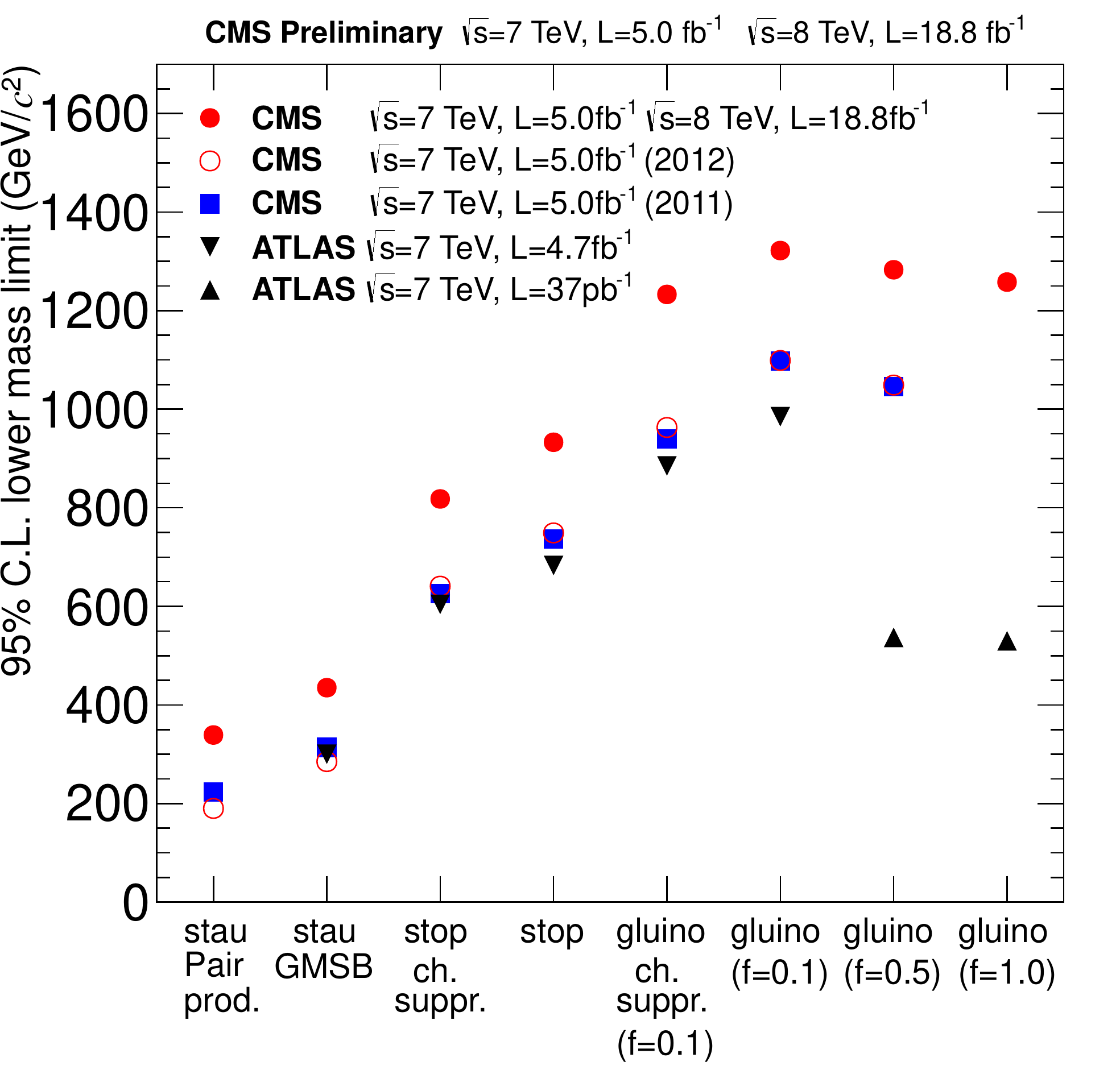}
  \hfill
  \includegraphics[clip=true, trim=0.0cm 0cm 1.0cm 0cm, width=0.45\linewidth]{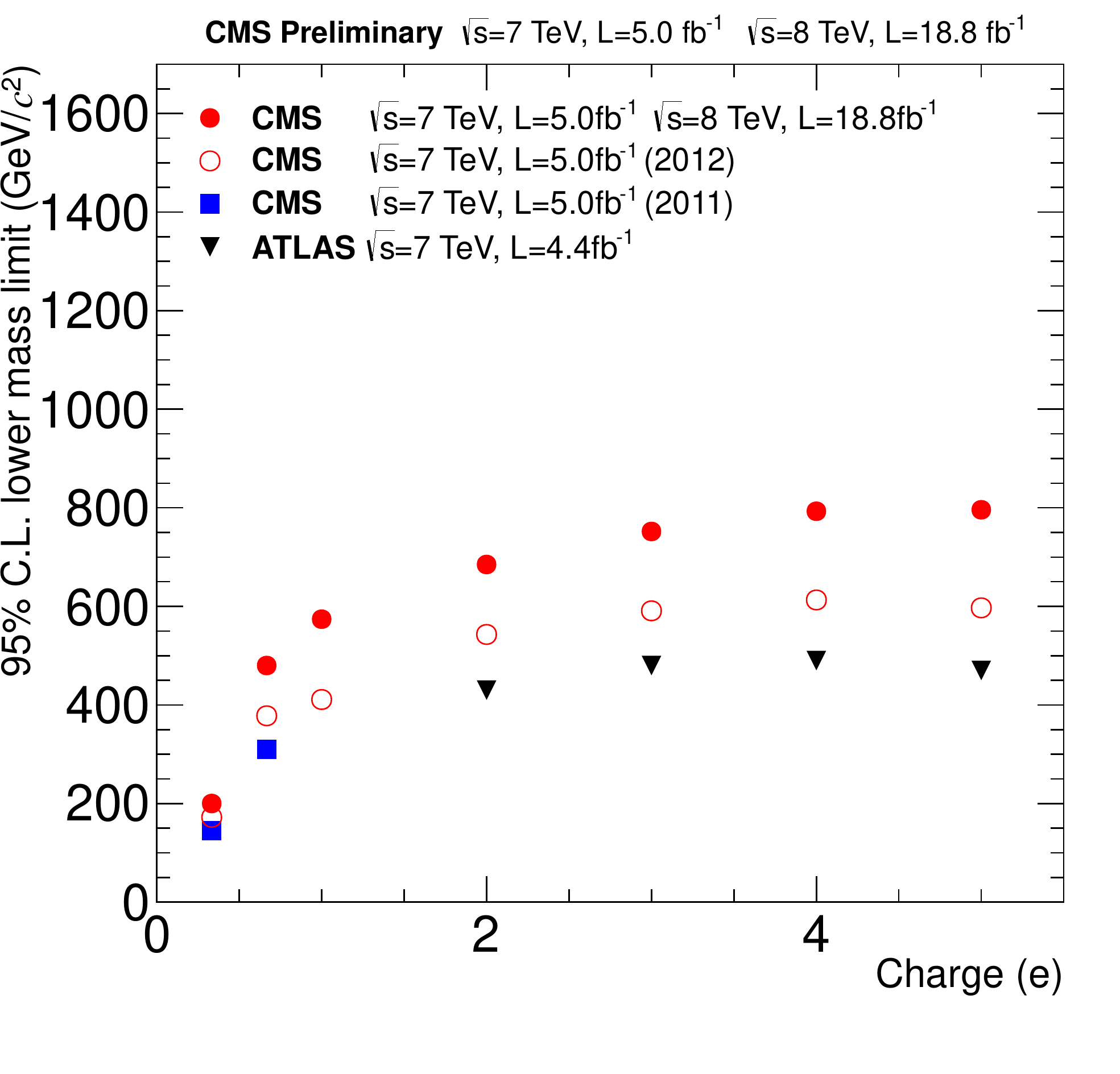}
 \end{center}
 \caption{Mass lower limits at 95\% CL on various models compared with previously published results. Left: The model type
is defined by the $x$ axis. Right: Mass limits versus electric charge.
   \label{fig:masslimits}}
\end{figure}

Many extensions to the SM predict particles with mass greater than 100 GeV and
particles with fractional charge or with charge greater than the
unit charge.
Such particles
could show very high or low energy loss as they travel through the detector (see the left plot of Fig.~\ref{figLimit}),
have a long time of flight,
or could switch between charged and neutral states while going through the detector. CMS has
performed a search~\cite{exo1226} for these particles with the
combined 2011 and 2012 pp data at 7 and 8 TeV, which totals 23.8~fb$^{-1}$. 
The search was done in five channels: using only the CMS inner silicon detectors (tracker), 
using the tracker plus the muon system, using only the muon system, using the tracker to search
for fractionally charged particles, and then using both systems to search for particles with
charge greater than the unit charge. The background estimates for these searches were derived from
data-driven methods. The systematic uncertainties on the signal acceptance range from 13--32\%,
while the uncertainty on the dominant background is about 20\%. The data is found to match the background
estimate within the uncertainties. Mass limits calculated at the 95\% confidence level (CL), shown in
Fig.~\ref{fig:masslimits}, are the most stringent to date for the signal models considered.
The limits range up to 1322 GeV on the gluino mass.

\section{Search for Narrow Resonances using the Dijet Mass Spectrum}
\label{sec:dijet}
\begin{figure}[hbtp]
  \begin{center}
  \includegraphics[clip=true, trim=0.0cm 0cm 0.0cm 0cm, width=0.45\linewidth]{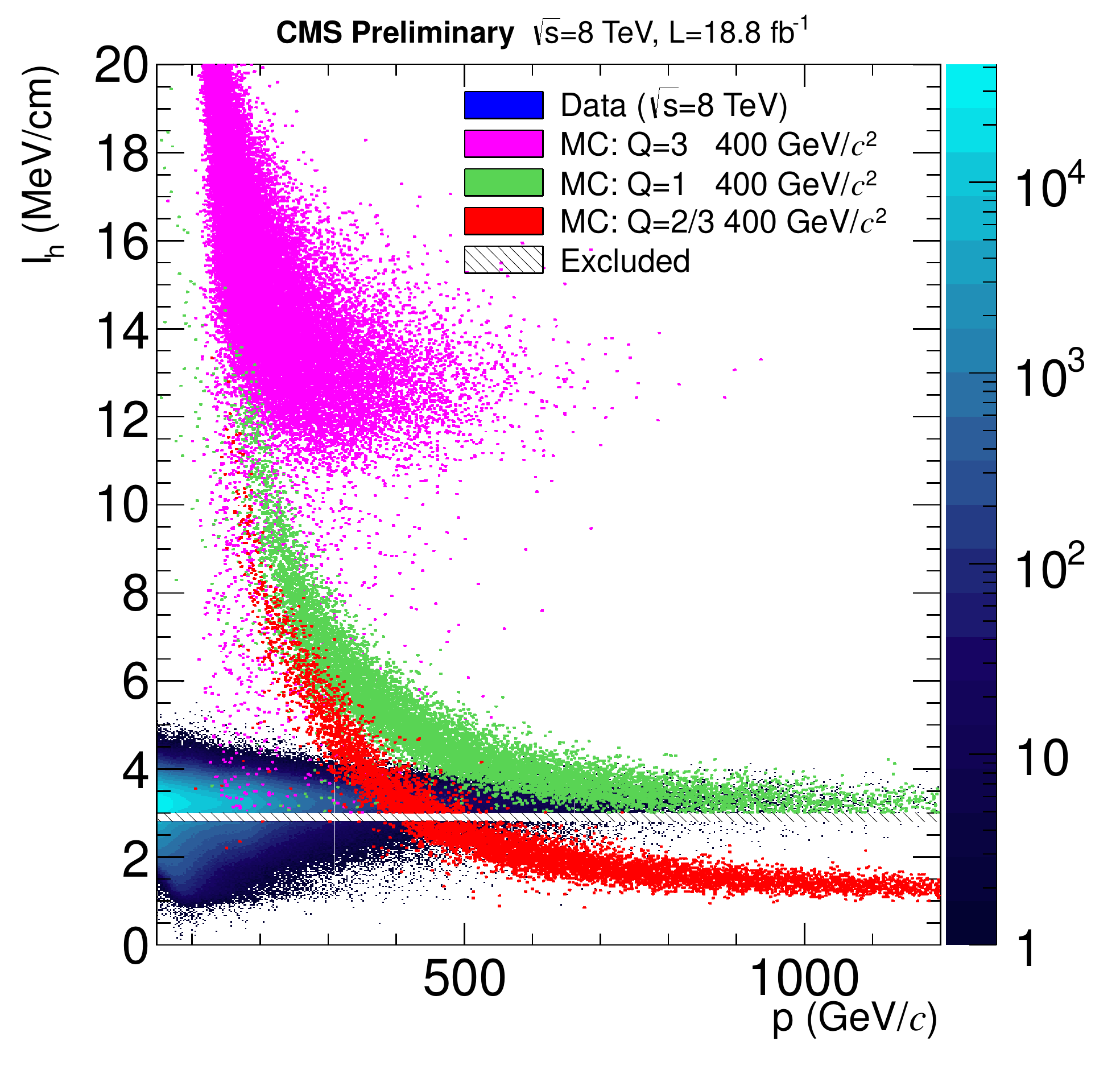}
  \hfill
    \includegraphics[width=0.45\linewidth]{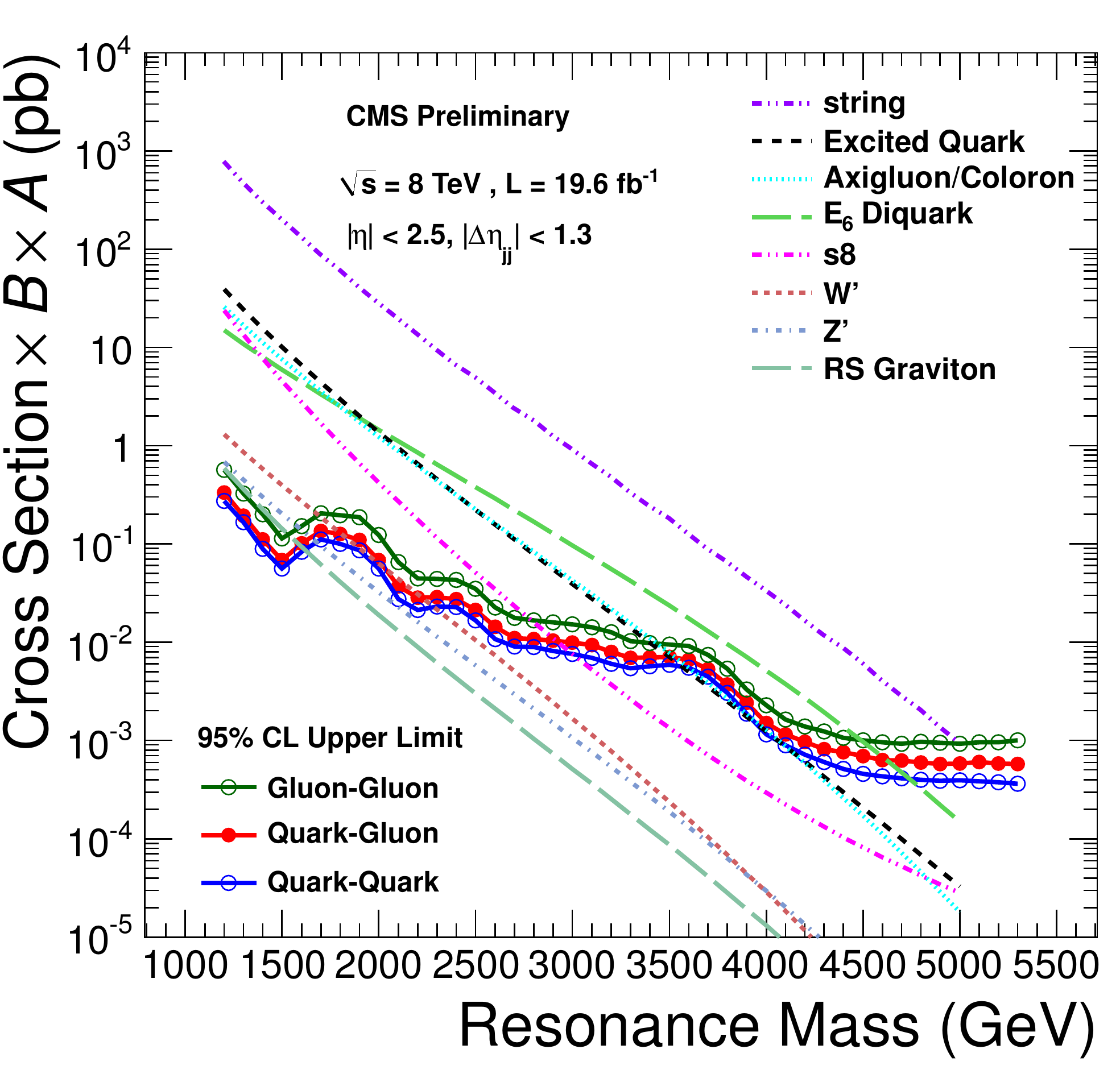}
    \caption{ On the left, for heavy stable charge particle candidates, the distribution of $I_h$, which is a measure of particle energy loss over distance, versus particle momentum for
   data and singly, fractionally, and multiply charged candidates.
   On the right, for the dijet search, 
the observed 95\% CL upper limits on $\sigma\times B\times A$
for dijet resonances of the type gluon-gluon, quark-gluon, and quark-quark,
compared to theoretical predictions for string resonances, $\mbox{E}_6$ diquarks,
excited quarks, axigluons,
colorons, s8 resonances, 
new gauge bosons $\mbox{W}^{\prime}$ and $\mbox{Z}^{\prime}$,
and RS gravitons. 
}
    \label{figLimit}
  \end{center}
\end{figure}

Many new physics models predict heavy resonances that couple to quarks and gluons and decay
to dijets. Some of these models include axigluons, color-octet colorons, excited quarks, 
Randall-Sundrum gravitons, scalar diquarks, string resonances, technicolor s8 resonances, W', and Z'.
CMS has performed a search for such resonances with the full 19.6~fb$^{-1}$ 2012 8 TeV pp dataset~\cite{exo1259}. This search employs the wide-jet technique which adds close sub-leading
jets to the two leading jets in each selected event. The background prediction comes from a
four-parameter fit to the data. The largest systematic uncertainty is the
jet energy resolution uncertainty, which is 10\%. The data matches the background estimate with
no excess or bumps observed on the smooth background. Mass limits are set on the eight signal
models, as shown on the right in Fig.~\ref{figLimit}, with the strongest limit being 5.1 TeV on the string
resonance mass.

\section{Search for Heavy Resonances Decaying into b$\bar{\textbf{b}}$ and bg Final States}
\begin{figure}[htbp]
  \begin{center}
    \includegraphics[width=0.32\textwidth]{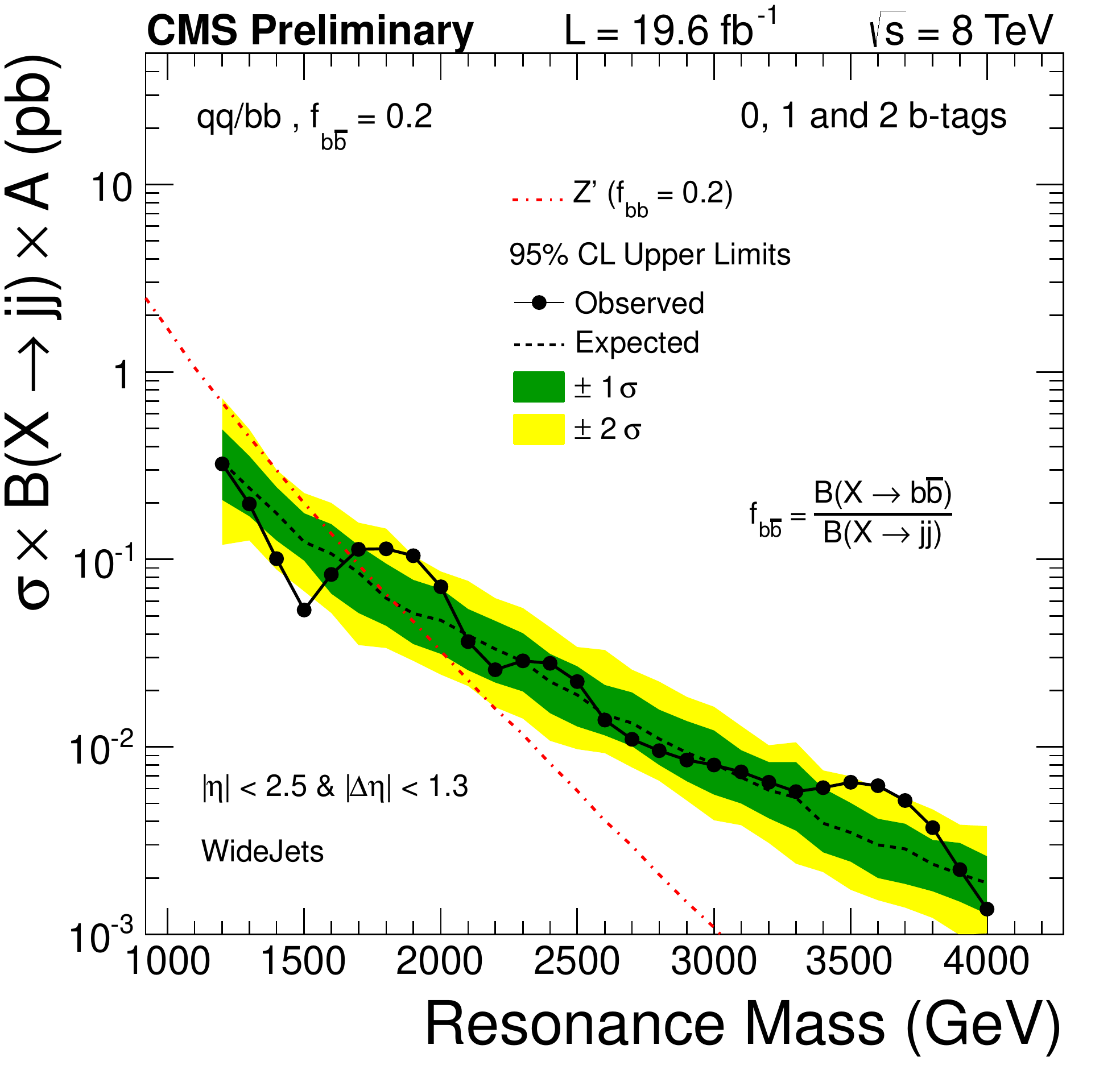}
    \includegraphics[width=0.32\textwidth]{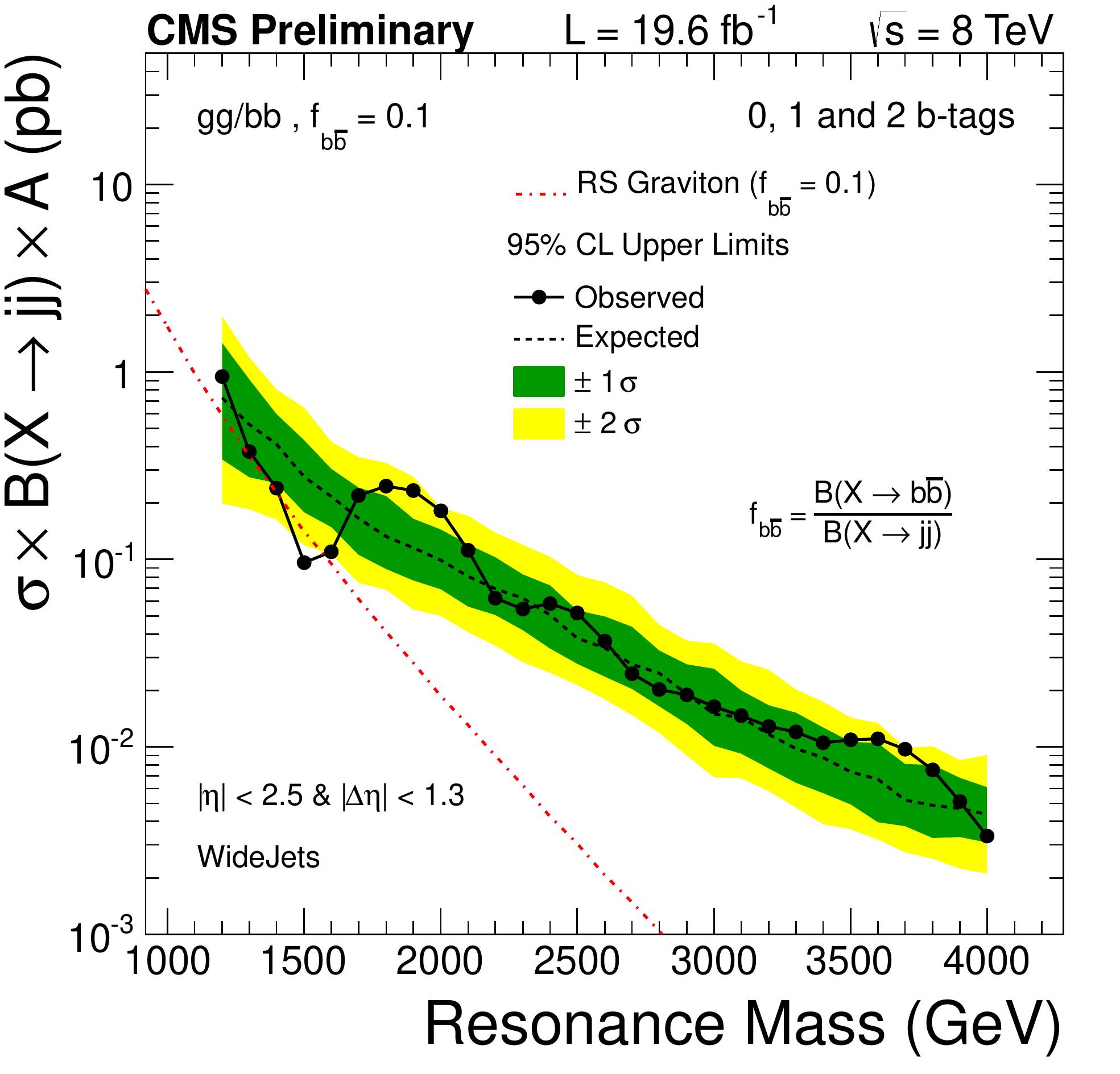}
    \includegraphics[width=0.32\textwidth]{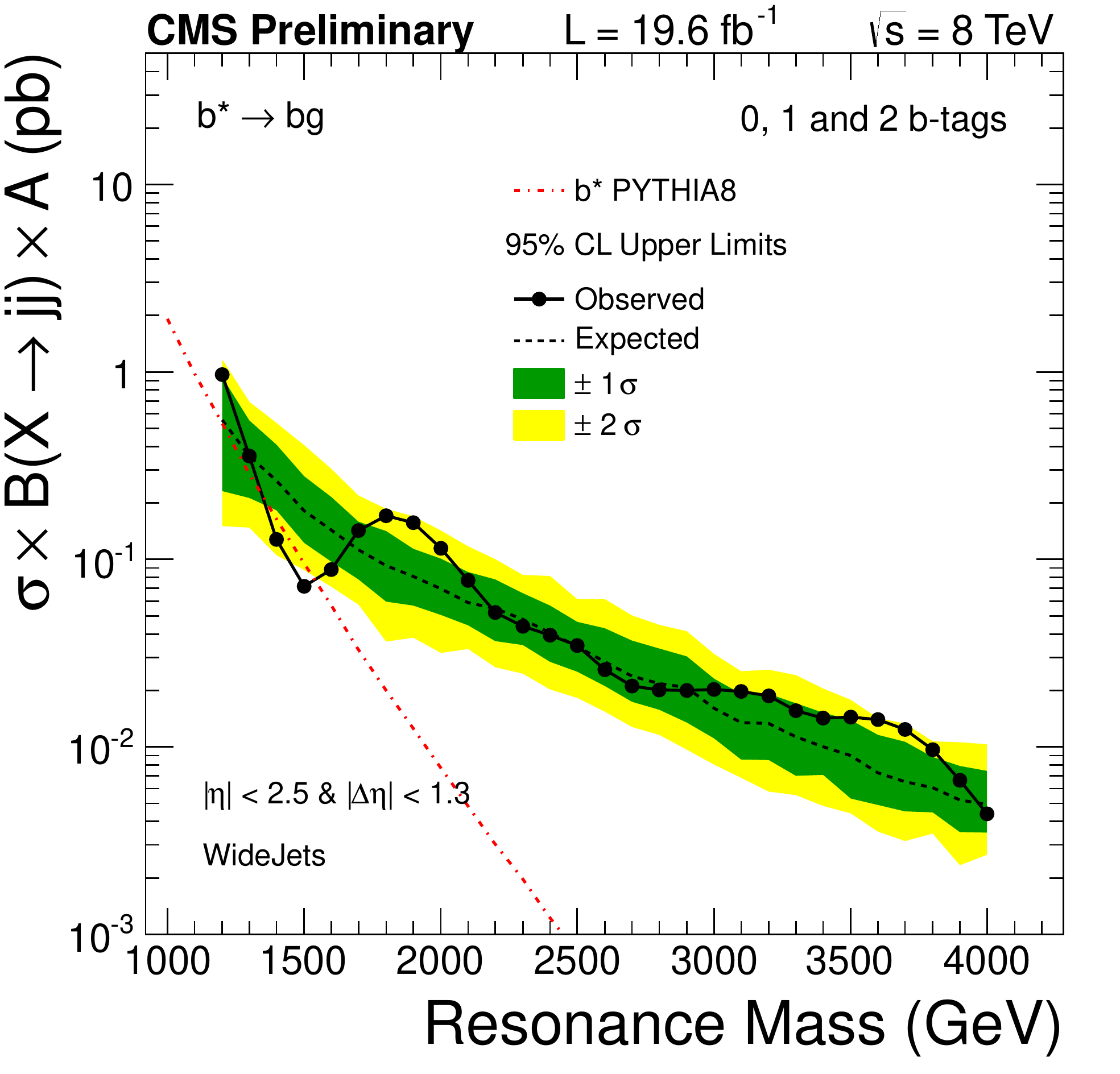}
    \caption{Combined observed and expected 95\% CL upper limits on $\sigma \times B \times A$ with systematic uncertainties
included for qq/bb resonances with $f_{\textrm{b} \bar{\textrm{b}}} = 0.2$ (left), gg/bb resonances with $f_{\textrm{b} \bar{\textrm{b}}} = 0.1$ (middle), and $\textrm{b}^{*} \rightarrow$ bg resonances (right). $f_{\textrm{b} \bar{\textrm{b}}}$ is the ratio
of the branching fraction of the resonance decaying to $\textrm{b} \bar{\textrm{b}}$ over the resonance decaying to all jets.
Theoretical cross sections for RS graviton, Z', and excited b quark are shown for comparison.}
    \label{fig:limits_obs_exp}
  \end{center}
\end{figure}

As a variation on the dijet search in Sec.~\ref{sec:dijet}, a b-jet tagging requirement can be
placed on the jets in order to reduce SM backgrounds and to make the search
sensitive to models that specifically produce b jets: excited b quarks, RS gravitons, and a
sequential SM Z'. CMS has performed such a search with the full 19.6~fb$^{-1}$ 2012 8 TeV pp dataset~\cite{exo1223}. Like the previous analysis, this one uses the wide-jet technique and
a background estimate from a four-parameter fit to the data. The jet energy resolution uncertainty
of 10\% is the largest systematic uncertainty. The data matches the  background estimate, and no excess is observed. The best mass limits to date are set on the three signal models, as shown in Fig.~\ref{fig:limits_obs_exp}, with the strongest limit being 1.7 TeV on the Z'
mass.

\section{Conclusion}
The CMS Collaboration has performed several searches for new physics using leptons and jets
with the 2012 8 TeV pp collision dataset. No significant deviations from the SM were observed.
New limits were set on many models, most being the best to date on these models. These limits
range up to 2.8 TeV on the mass of a $\textrm{W}_{\textrm{R}}$, up to 1.1 TeV on the mass of
scalar leptoquarks, up to 1.3 TeV on high-mass supersymmetric particles with anomalous charges,
up to 5.1 TeV on dijet string resonances, and up to 1.7 TeV on a Z'. CMS is continuing its searches
for new physics, and more results will be coming out soon.

\bibliography{auto_generated}   

\providecommand{\href}[2]{#2}\begingroup\raggedright\begin{thebibliography}{1}%
\makeatletter
\providecommand{\hrefCMSnoop }[0]{\@secondoftwo}%
\makeatother
\providecommand{\doi}{\texttt{doi:}\begingroup \urlstyle{tt}\Url}

\bibitem{cmsdet}
\hrefCMSnoop {} {{ CMS} Collaboration, ``\emph{The {CMS} experiment at the
  {CERN} {LHC}}'',} \textit{ JINST} \textbf{ 3} (2008) S08004,
\href{http://dx.doi.org/10.1088/1748-0221/3/08/S08004}{\doi{10.1088/1748-0221/3/08/S08004}}.

\bibitem{exo1217}
\href {http://cdsweb.cern.ch/record/1460445} {{ CMS} Collaboration,
  ``\emph{Search for a heavy neutrino and right-handed W of the left-right
  symmetric model in pp collisions at sqrt(s) = 8 TeV}'',} \textit{ CMS Physics
  Analysis Summary} \textbf{ CMS PAS EXO-12-017} (2012).

\bibitem{exo1242}
\href {http://cdsweb.cern.ch/record/1542374} {{ CMS} Collaboration,
  ``\emph{Search for pair production of second-generation scalar leptoquarks in
  pp collisions at sqrt(s) = 8 TeV with the CMS detector}'',} \textit{ CMS
  Physics Analysis Summary} \textbf{ CMS PAS EXO-12-042} (2012).

\bibitem{exo1226}
\href {http://cdsweb.cern.ch/record/1529897} {{ CMS} Collaboration,
  ``\emph{Search for Long-lived Charged Particles in pp collisions at sqrt(s) =
  7 and 8 TeV}'',} \textit{ CMS Physics Analysis Summary} \textbf{ CMS PAS
  EXO-12-026} (2013).

\bibitem{exo1259}
\href {http://cdsweb.cern.ch/record/1519066} {{ CMS} Collaboration,
  ``\emph{Search for Narrow Resonances using the Dijet Mass Spectrum with 19.6
  $fb^{-1}$ of pp Collisions at sqrt(s) = 8 TeV}'',} \textit{ CMS Physics
  Analysis Summary} \textbf{ CMS PAS EXO-12-059} (2013).

\bibitem{exo1223}
\href {http://cdsweb.cern.ch/record/1542405} {{ CMS} Collaboration,
  ``\emph{Search for Heavy Resonances Decaying into bbbar and bg Final States
  in pp Collisions at sqrt(s) = 8 TeV}'',} \textit{ CMS Physics Analysis
  Summary} \textbf{ CMS PAS EXO-12-023} (2013).

\end{thebibliography}\endgroup

\end{document}